\documentclass[conference]{IEEEtran}
\IEEEoverridecommandlockouts

\usepackage{tikz}
\usepackage{lipsum} 

\makeatletter
\def\ps@IEEEtitlepagestyle{%
  \def\@oddfoot{}%
  \def\@evenfoot{}%
  \def\@oddhead{%
    \parbox[b]{\textwidth}{\centering
    Author copy of paper published at \textit{2025 IEEE International Conference on Edge Computing and Communications (EDGE)}\\
    \copyright~2025 IEEE\vspace{1em}}
  }%
  \def\@evenhead{\@oddhead}%
}
\makeatother

\usepackage{cite}
\usepackage{amsmath,amssymb,amsfonts}
\usepackage{array,multirow}
\usepackage{algorithmic}
\usepackage{graphicx}
\usepackage{textcomp}
\usepackage{xcolor}
\usepackage{soul}
\usepackage{pythonhighlight}
\def\BibTeX{{\rm B\kern-.05em{\sc i\kern-.025em b}\kern-.08em
    T\kern-.1667em\lower.7ex\hbox{E}\kern-.125emX}}

\usepackage{listings}    
\newcommand\YAMLcolonstyle{\color{charcoal}\mdseries\footnotesize}
\newcommand\YAMLkeystyle{\color{black}\bfseries\footnotesize}
\newcommand\YAMLvaluestyle{\color{arsenic}\mdseries\footnotesize}

\makeatletter
\usepackage{float}
\usepackage{syntax}
\usepackage{newfloat}
\usepackage{textcomp}
\usepackage{subfig}
\usepackage{tabularx}
\usepackage{balance}
\usepackage[para]{footmisc}
\usepackage{multirow}
\usepackage{listings}
\definecolor{charcoal}{rgb}{0.21, 0.27, 0.31}
\definecolor{arsenic}{rgb}{0.23, 0.27, 0.29}
\newcommand\language@yaml{yaml}

\expandafter\expandafter\expandafter\lstdefinelanguage
\expandafter{\language@yaml}
{
  keywords={true,false,null,y,n},
  keywordstyle=\color{darkgray}\bfseries,
  basicstyle=\YAMLkeystyle,
  sensitive=false,
  comment=[l]{\#},
  morecomment=[s]{/*}{*/},
  commentstyle=\color{purple}\ttfamily,
  stringstyle=\YAMLvaluestyle\ttfamily,
  moredelim=[l][\color{orange}]{\&},
  moredelim=[l][\color{magenta}]{*},
  moredelim=**[il][\YAMLcolonstyle{:}\YAMLvaluestyle]{:},   
  morestring=[b]',
  morestring=[b]",
  literate =    {---}{{\ProcessThreeDashes}}3
                {>}{{\textcolor{red}\textgreater}}1     
                {|}{{\textcolor{red}\textbar}}1 
                {\ -\ }{{\mdseries\ -\ }}3,
}

\lst@AddToHook{EveryLine}{\ifx\lst@language\language@yaml\YAMLkeystyle\fi}
\makeatother

\newcommand\ProcessThreeDashes{\llap{\color{cyan}\mdseries-{-}-}}
\usepackage{cleveref}
\usepackage{paralist}
\usepackage{url}
\makeatletter
\newcommand{\linebreakand}{%
  \end{@IEEEauthorhalign}
  \hfill\mbox{}\par
  \mbox{}\hfill\begin{@IEEEauthorhalign}
}
\makeatother

\usepackage[pscoord]{eso-pic}

\begin{document}
\title{Automating Multi-Tenancy Performance Evaluation on Edge Compute Nodes}
 \author{
 \IEEEauthorblockN{Joanna Georgiou, Moysis Symeonides, George Pallis, Marios D. Dikaiakos}
 \IEEEauthorblockA{Department of Computer Science, University of Cyprus\\
 Email: \{jgeorg02, msymeo03, gpallis, mdd\}@ucy.ac.cy}
 }
\captionsetup{font=footnotesize} 

\maketitle

\begin{abstract}
Edge Computing emerges as a promising alternative of Cloud Computing, with scalable compute resources and services deployed in the path between IoT devices and Cloud. 
Since virtualization techniques can be applied on Edge compute nodes, administrators can share their Edge infrastructures among multiple users, providing the so-called multi-tenancy.
Even though multi-tenancy is unavoidable, it raises concerns about security and performance degradation due to resource contention in Edge Computing. For that, administrators need to deploy services with non-antagonizing profiles and explore workload co-location scenarios to enhance performance and energy consumption. Achieving this, however, requires extensive configuration, deployment, iterative testing, and analysis, an effort-intensive and time-consuming process. To address this challenge, we introduce an auto-benchmarking framework designed to streamline the analysis of multi-tenancy performance in Edge environments. Our framework includes a built-in monitoring stack and integrates with widely used benchmarking workloads, such as streaming analytics, database operations, machine learning applications, and component-based stress testing. We perform a case-driven analysis and provide valuable insights into the impact of multi-tenancy on Edge environments with different hardware configurations and diverse workloads.
Finally, the implementation of our framework, along with the containerized workloads used for experimentation, is publicly available.
\end{abstract}

 \vspace*{-0.7\baselineskip}
\section{Introduction}
The rapid expansion of interconnected embedded devices, collectively known as the Internet of Things (IoT), is transforming our daily lives. The number of IoT devices is expected to reach 500 billion by 2030~\cite{iotforecast}, with each device generating vast amounts of data that require analysis.
However, processing data 
on an IoT device is often impractical due to its limited computational capacity and reliability~\cite{Symeonides2020}. Meanwhile, continuously offloading data to centralized cloud infrastructures introduces challenges, including bandwidth constraints~\cite{Weisong2016}.

To bridge this gap, Multi-access Edge Computing (MEC) has emerged as a standardized architecture that, like traditional edge computing, brings processing closer to the IoT devices to enhance efficiency and reduce latency. What distinguishes MEC, however, is its tight integration with the Radio Access Network (RAN), enabling more effective utilization of network resources and further optimizing performance.~\cite{porambage2018survey}.
Building on top of this, the 5th/6th generations of mobile networks (5G and 6G) advance the concept of Network Slicing, enabling the creation of multiple isolated virtual networks over a shared physical infrastructure. This allows operators to support multi-tenancy by leasing tailored ``slices" of their networks, similarly to how Cloud providers rent out compute~resources~\cite{wijethilaka2021survey}.With multi-tenancy, high-speed and performance applications like autonomous driving, drone-based surveillance,
agentic workflows and LLM integration can be supported more effectively, as these types of workloads often require multiple machine learning, deep learning and LLM tasks to be executed in parallel~\cite{zobaed2022, Subedi2021, alla2024sound,krasovitskiy2025}.

Multi-tenancy refers to a system's ability to simultaneously serve multiple tenants (users or applications), while ensuring strict isolation of their data and resources~\cite{sharmaContainersVsVms2016}. 
A key enabler technology for multi-tenancy on Edge nodes is a lightweight virtualization (containerization) which bundles services into artifacts and runs them as isolated processes with negligible computational overhead~\cite{Morabito2017}. 
Generally, service consolidation reduces the operational and maintenance costs of the infrastructure, while it also decreases energy consumption~\cite{Premsankar2022}. 

Unfortunately, even in cloud infrastructures, multi-tenancy raises concerns about security and performance deterioration of co-located services. 
Workloads that share the same resources may suffer quality-of-service degradation due to contention for CPU, memory, and network bandwidth. The constrained and heterogeneous nature of Edge devices, along with the diverse profiles of the deployed services, worsens these challenges.

For instance, running a set of services on an edge server can deliver high performance, but executing the same services on a resource-constrained device, such as a Raspberry Pi, may significantly increase service latency. 
This can be problematic for applications that require real-time responsiveness while performing multiple deep learning tasks simultaneously, such as object detection and face recognition for visually impaired users~\cite{zobaed2022}.
To maintain efficiency, edge infrastructure administrators and cluster orchestrators must co-locate services with compatible resource profiles on the same machine. This enables them to (i) assess whether local resources can sustain the workload and (ii) analyze trade-offs associated with multi-tenancy. These considerations are central to informed orchestration decisions.
Therefore, it is essential to identify \textit{which deployment scenarios or configurations enhance the performance of new or existing services and how to achieve
objectives, such as minimizing energy consumption.}

Addressing this challenge requires the repeated deployment of diverse workloads across a range of configurations, to explore the full spectrum of co-location scenarios, and the continuous collection of performance metrics and infrastructure utilization data, followed by rigorous analysis to identify optimal deployment strategies~\cite{georgiou2022benchpilot}. However, as the number of potential configurations grows exponentially with the number of target workloads~\cite{Rafiuzzaman2024}, the process quickly becomes intractable. 
Even though there are many benchmarks that prototype workloads like big data streaming~\cite{yahooStreaming2016benchmarking} and machine learning (ML) inference~\cite{reddi2020mlperf}, very few studies~\cite{subedi2021ai, zobaed2022} actually examine the effects of multi-tenancy in edge environments. Moreover, although tools for automating repeatable performance evaluations are available~\cite{georgiou2022benchpilot, jansen2023continuum}, they lack support for executing concurrent workloads.
This paper addresses the critical gap: the absence of a low-cost, scalable and reproducible approach for evaluating performance trade-offs introduced by workload co-location in real-world edge environments. The main contributions of this paper are:
\begin{itemize}
 \item An open-source benchmarking framework~\cite{github}
, that facilitates the automated deployment of benchmarking experiments and ensures reproducible performance evaluation of co-located workloads on top of real edge infrastructures. It incorporates comprehensive monitoring for both performance and infrastructure metrics, supports detailed post-experiment analysis, and addresses common challenges in performance evaluation. Our framework leverages cloud-native technologies, including continuous deployment and containerization, to streamline benchmark configuration and deployment, whilst abstracting the hardware heterogeneity across the edge continuum.

 \item A set of publicly accessible containerized workloads~\cite{dockerhub}
 , integrated with our framework to evaluate the performance of edge nodes. These workloads are derived from widely adopted benchmarks~\cite{yahooStreaming2016benchmarking, reddi2020mlperf, cooper2010benchmarking} and component-based stressors, and each has been modified to support parameterization and easy configuration.
 As the workloads are independent of the framework, researchers can utilize, extend, or adapt them to their requirements.

 \item A comprehensive, use-case-driven analysis of workload co-location on various edge compute nodes (from Raspberry Pi to edge Servers).    
Our analysis allows users to assess:  
(i)~ML service performance across various compute nodes, showing that larger Edge servers offer better performance at higher power costs, while smaller devices like Raspberry Pi are more energy-efficient but slower;
(ii)~the impact of workload co-location, revealing that memory-intensive workloads increase cold start duration of ML services by 1.5x, and CPU or disk I/O stressors reduce CPU availability, degrading performance;
(iii)~the effects of different co-location configurations, comparing ML inference in two scenarios: (a) processing images locally vs streaming them, with performance varying by up to 3x, and (b) using different execution backends, where ONNX (Open Neural Network Exchange) outperforms NCNN (inference framework optimized for mobile platforms) by 9x and TensorFlow by 24x in throughput.
\end{itemize}

The remainder of the paper is structured as follows: Sec.\ref{sec:motivation} outlines the motivating scenario; Sec.\ref{sec:related-work} reviews related work; Sec.\ref{sec:framework} introduces our framework; Sec.\ref{sec:workloads} details the workloads and metrics used for evaluating multi-tenancy on Edge devices in Sec.\ref{sec:experimentation}; and Sec.\ref{sec:conclusion} concludes the paper.

\section{Motivating Example}
\label{sec:motivation}

To illustrate the applicability of our framework, consider a scenario where an edge infrastructure provider, such as a mobile operator, serves multiple clients deploying applications on a shared Edge environment. The infrastructure consists of various heterogeneous devices, from high-performance edge servers to resource-constrained micro-edge devices like single-board computers. To meet client demands, the provider enables containerized services to run in parallel across these nodes while ensuring a specified Quality of Service (QoS).


Co-locating applications, even on constrained devices, is often more practical despite potential performance degradation from resource contention, as dedicating a node to each workload is usually impractical due to power, cost, and resource limitations.~\cite{wang2020dyverse}.
Operators, researchers, and system administrators who want to study multi-tenancy effects, optimize deployments, and benchmark their infrastructure must carefully assess the impact and trade-offs of workload co-location to determine optimal deployment strategies.
Factors like energy efficiency must also be considered, requiring extensive experimentation to identify configurations that minimize energy consumption while maintaining performance~\cite{trihinas2023energy}. Hence, researchers must either develop custom workloads, an effort requiring significant time and expertise, or use existing workloads, which still involve extensive software and hardware setup, dependency management, and monitoring configurations. 
Then they must execute a benchmarking pipeline, which traditionally involves manually deploying various workloads, monitoring resources, and analyzing service performance. This process on its own is time-consuming and shifts the focus away from the user's objective, which is the system's performance evaluation~\cite{georgiou2022benchpilot}. 
However, in the case of assessing multi-tenancy, this task becomes even more challenging, as it requires testing all workload combinations with diverse configurations. 
Additionally, as the number of target workloads and configurations increases, the number of scenario combinations grows exponentially, 
lengthening the process even further.

To address these challenges, our open-source framework streamlines the benchmarking of co-located services, enabling users to assess infrastructure performance, fine-tune service parameters, and analyze co-location trade-offs, such as identifying antagonistic workload metrics, without the overhead of managing complex setup and configuration tasks.
The methodology automates (i) the installation and configuration of essential software components, including containers and monitoring agents on edge nodes, and (ii) the deployment of both isolated and co-located, parameterizable services (e.g., database workloads, ML inference) across diverse scenarios.

\section{Related Work}
\label{sec:related-work}

Analyzing interference between applications in data centers is a well-known issue, requiring workload analysis through infrastructure metrics. Towards that, Ilager et al.\cite{ilager2023} conducted a data-driven study on cloud datacenter workloads, energy, and thermal characteristics using nine months of machine-level metrics, developing predictive models for efficiency planning. Similarly,\cite{wenyan2019} analyzed workload interference in cloud providers (Google, Alibaba), quantifying co-location effects at micro-architecture and app-level. In~\cite{prateek2016}, virtualization techniques (containers, VMs) were assessed, showing higher interference in containers, while an experimental study in~\cite{nandan2018} examined performance interference in containerized microservices, measuring the effects of co-located services. However, \textit{these studies focus solely on workload co-location in cloud environments, 
omitting potential issues in MEC~deployments.}

Recognizing the unique constraints of edge environments, edge co-location research has primarily focused on the deployment of ML inference workloads~\cite{zobaed2022, Subedi2021}. For example, Edge-MultiAI enhances the multitenancy of ML inference while meeting latency and accuracy goals~\cite{zobaed2022}, while~\cite{Subedi2021} introduces concurrent Deep Learning model inference with dynamic placement, improving throughput on Jetson TX2 using various ML frameworks. However, beyond ML-specific workloads, only a few systems consider general-purpose co-location on the edge, mostly from a scheduling perspective. PolarisProfiler~\cite{Morichetta2023} optimizes resource management through metadata-based profiling, while Edge Federation~\cite{awada2020} introduces a dependency-aware scheduler for federated containerized workloads. Mathematical approaches like Co-Approximator~\cite{Rafiuzzaman2024} estimate co-location performance but lack automation and scalability considerations. Similarly, efforts to optimize network co-location (cellular base stations and MECs), such as ColoMEC~\cite{Minh2020}, focus on orchestration but do not address general-purpose workload evaluation. Consequently, \textit{the gap in automating the performance evaluation of co-located workloads on edge infrastructures still remains}.

To address the evaluation of edge deployments, experimentation tools are widely used by Edge Computing engineers to compare frameworks, applications, and infrastructures. Fog and Edge emulators~\cite{Symeonides2020, Ruben2017, Jonathan2023, Matthijs2023, naman2023mecbench} facilitate topology emulation, application deployment, and metric extraction but do not support automated deployment on physical infrastructures. Although some frameworks attempt to automate evaluation, their focus remains limited. For instance, Plug and Play Bench (PAPB)\cite{Ceesay2017} enables big data benchmarking using containerized execution, while Frisbee\cite{Nikolaidis2021} performs chaos engineering on Kubernetes-deployed applications with a declarative approach to failure injection. Similarly, BenchPilot~\cite{georgiou2022benchpilot}, although it provides declarative experimentation descriptions and benchmarking functionalities, supports only a single type of workload and, more importantly, does not support co-location. While these systems demonstrate promising advancements, they \textit{
do not provide mechanisms for automating the performance evaluation and analysis of co-located workloads across diverse edge environments.}

In summary, although several existing works support benchmarking or workload orchestration in cloud/edge settings, they either lack the capability to evaluate multi-tenancy, especially on real edge nodes, or focus on a specific type of workloads. Our work distinguishes itself by enabling fully automated deployment and in-depth analysis of co-located workloads. This empowers operators and researchers to explore interference patterns and deployment trade-offs across real-world, diverse, edge environments in a reproducible manner.

\section{Benchmarking Framework}
\label{sec:framework}

\subsection{System Overview}
\begin{figure}[t]
  \centering
     \includegraphics[width=0.48\textwidth]{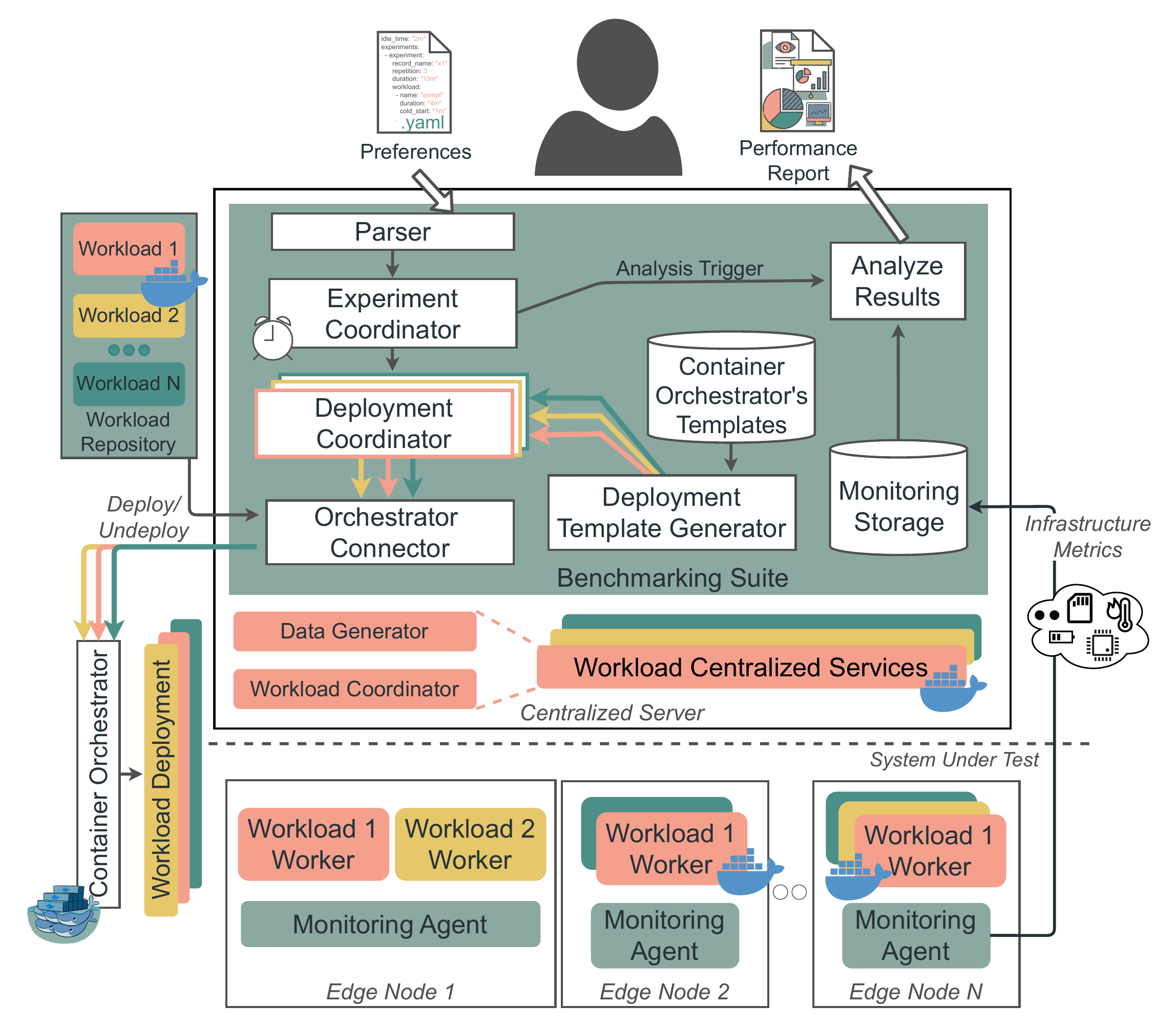}
           \vspace*{-.5\baselineskip}

    \caption{The Overview of the Auto-benchmarking Framework}
    
  \label{fig:framework-overview}
             \vspace*{-1.5\baselineskip}

\end{figure}

To address the complexity of evaluating and the deployment of co-located workloads,
our system initially undergoes a bootstrapping phase to automatically install and configure all required dependencies across the evaluation nodes if necessary. 
This process ensures that all nodes are properly set up based on the user's configuration file that defines nodes' roles and access credentials (see Section~\ref{sec:bootstrapping}).
Once the bootstrapping phase is complete, the user can proceed with submitting experiment descriptions, as depicted in Fig.~\ref{fig:framework-overview}.

Specifically, the experiment description includes a list of co-located workloads, their configurations, execution duration, potential starting delays, and more, all structured according to the framework's high-level YAML model (see Section~\ref{sec:model}).
Having a submitted description, the \textit{Parser} is responsible for parsing it and evaluating whether the description of the experiments is syntactically correct and valid.
If the description is correct, it is propagated to the~\textit{Experiment Coordinator}. This module translates the given information into a list of experiment objects and organizes their execution.

For each experiment object, the~\textit{Deployment Coordinator} will be spawned to create the workload(s) configurations.
Each workload supported by the framework has a corresponding \textit{Container Orchestrator Template} that defines the required services, a set of default parameters (which can be overridden), and their respective container images. The \textit{Deployment Template Generator} retrieves the appropriate template, populates it based on the user's submitted variables, generates the final workload description, and forwards it to the Deployment Coordinator.  
Once all of the experiment and workload objects have been prepared, the~\textit{Orchestrator Connector} creates deployment descriptions compatible with the user's desired container orchestrator. 
The orchestrator is responsible for managing the deployment and networking of containerized services and is materialized by frameworks like Docker Swarm.

After the benchmarking process setup is completed, the \textit{Experiment Coordinator} will coordinate additional actions: (i)~recording the starting and ending timestamps of the experiment, (ii)~giving instruction to trigger workloads that were meant to start ~\textit{delayed}, (iii)~monitoring all workloads to ensure they are healthy, up, and running, and (iv)~upon completion of an experiment, either repeating it as many times as defined by the user or starting the next experiment. To accomplish this, the \textit{Experiment Coordinator} maintains constant communication with the \textit{Deployment Coordinator}, which is responsible for invoking, monitoring, and stopping each deployment (workload).
Once all experiments are complete, the user can retrieve the collected metrics and focus on the analysis. 
Specifically, our framework collects metrics--such as CPU usage, memory consumption, network traffic, and energy consumption--for both the underlying hardware and each deployed service. Additionally, it extracts workload-specific metrics, like latency or throughput, and stores all data as CSV files.



\subsection{Bootstrapping Process}
\label{sec:bootstrapping}

To streamline the benchmarking setup, our framework automates the installation, configuration, and deployment of all required software and dependencies, including the monitoring stack. Before benchmarking begins, it connects to each node based on a bootstrapping configuration where users define the cluster setup, specifying the manager and worker nodes along with their access credentials.
Description~\ref{boostrapping-model} presents an example configuration file, where users specify: (i) the \textit{manager} node, responsible for centralized services, such as data generators, and (ii) the \textit{worker} nodes to be evaluated. Each node entry includes an accessible IP (reachable from the node where the framework is set up), hostname, and login credentials, supporting authentication via (i)~username-password or (ii)~username-SSH key.
This file only needs to be set up once unless the user wishes to introduce new nodes. 
When the user initiates a benchmarking process, the framework utilizes this information to prepare the nodes for the evaluation, by installing all necessary software and dependencies, and then proceed with the benchmarking process.

\begin{figure}[t]
\begin{lstlisting}[language=Python, caption={Example of the Bootstraping Configuration}, label={boostrapping-model}]
cluster:
  manager:
      ip: "0.0.0.0" # manager's IP
      hostname: "manager"
      ssh_key_path: "/conf/ssh_keys/ssh_key.pem" 
  nodes: # system under test
    - ip: "10.10.10.10" # using password
      hostname: "raspberrypi"
      username: "pi"
      password: "raspberrypi"
   ...
\end{lstlisting}
\vspace*{-2\baselineskip}
\end{figure}


\subsection{Experimentation Modeling}
\label{sec:model}

Our framework offers users a YAML-based representation for defining comprehensive and repeatable experiment descriptions at a higher abstraction level. Users can specify a collection of experiment elements within the \textit{experiments} section of the YAML file. Each experiment is identified by a \textit{record\_name}, which serves as the name of the log file that the system will generate for the experiment. Then, users specify the number of times the experiment should be executed~(\textit{repetition}) and, its execution \textit{duration}. Once these parameters are specified, they define a set of \textit{workloads} that the system will deploy concurrently. Description~\ref{experiment-model} presents a representative example of our modeling configuration, where two distinct workloads are defined: a database and a streaming analytics workload. Focusing on the first workload, users declare its \textit{name} according to the framework's supported workloads. Then, they designate the set of nodes (\textit{cluster}) where the workload will be deployed and provide the relevant workload parameters (e.g., \textit{db: "mongodb"}). Depending on the workload type, users can also specify additional parameters, such as the data generation rate (e.g., \textit{tuples_per_second}), the underlying system type (e.g., \textit{"storm"}, \textit{"mongodb"}, etc.), and other configuration settings. Users can also introduce a delay (\textit{shift}) for one or more workloads, allowing them to start a few seconds or minutes after the experiment begins. Once the experiment definitions are completed, they can configure additional parameters, such as the idle duration between experiments (\textit{idle_between_experiments}) and the chosen connector for managing the cluster (\textit{orchestrator}, e.g., \textit{"docker swarm"}). It is important to highlight that, since our framework operates directly on the underlying 
infrastructure without imposing additional constraints 
on system resources, users do not need to specify any limitations, such as memory and network~settings.


\subsection{Containerization of Workloads}

Containerization is a lightweight virtualization technique that abstracts hardware, enabling efficient application execution with minimal performance overhead \cite{Morabito2017}.
To achieve an efficient and dynamic workload deployment, our framework utilizes Docker as its containerization engine. 
In Docker, users define services using Dockerfiles, which specify configurations, dependencies, and environment variables before being built into images and deployed as running containers. By leveraging Docker's inheritance mechanism, we introduce a base image, containing all essential dependencies, to ensure consistency across workloads.
To support diverse edge infrastructures, including x64 and ARM architectures, our framework generates multi-architecture Docker images, automatically compiling them for the target hardware. For seamless distribution, we publish all image artifacts on DockerHub~\cite{dockerhub}, making them accessible to the research and industry communities.
Finally, containerizing the workloads enables our system to seamlessly integrate with various Docker orchestrators, such as Docker Swarm or Kubernetes, without requiring any modifications to its core implementation components.

\subsection{Infrastructure Monitoring Stack \& Metrics}
To monitor workloads and the system under test, our framework employs a monitoring stack that collects infrastructure metrics such as CPU and memory usage.
This is enabled by deploying a containerized Netdata~\cite{netdata2025} agent on each node to gather metrics non-intrusively via multiple probes targeting specific sub-components (e.g., cgroup, OS). The agent exposes an API that a centralized monitoring server uses to periodically collect and store data. 
In our testbed, we expose two probes: one for a Meross Wi-Fi Smart Plug~\cite{meross2025} 
for IoT devices, and another for a Smart Power Distribution Unit (SPDU) that uses SNMP to retrieve server power consumption. All metrics are collected at 5s intervals and stored in Prometheus~\cite{prometheus2025}, a widely used open-source monitoring server.

The key metrics used in our evaluation are: (i)~average \textit{CPU Utilization} (excluding idle time, in \%), as high usage can lead to performance issues and crashes; (ii)~total \textit{Memory Usage} (in MiB), to ensure sufficient memory for system stability; (iii)~total \textit{Disk I/O} (in KiB), which affects data-intensive task performance; (iv)~total \textit{Network I/O} (in bytes), as high usage may cause delays; (v)~average \textit{Power Consumption} (in watts), a key cost and environmental factor; and (vi)~average \textit{CPU Temperature} (in~$^\circ$C), crucial for preventing hardware damage. All metrics are collected at the system level, while CPU, memory, disk, and network are collected on service level.


\begin{figure}[t]
\begin{lstlisting}[language=Python, caption={Example of the Experiment Model}, label={experiment-model}]
 experiments:
   - experiment:
       record_name: "streaming_with_db"
       repetition: 2
       duration: "20m"
       workloads:
         - name: "database"
           cluster: [ "rpi", "small_server" ] 
           parameters:
             db: "mongodb"
         - name: "marketing-campaign"
           cluster: [ "rpi", "small_server" ]
           parameters:
             engine: "storm"
             enging_parameters:
                 tuples_per_second: 1000
                 capacity_per_window: 10
           shift: 5m
   - experiment: 
            ...
 idle_between_experiments: "2m"
 orchestrator: "docker swarm"
\end{lstlisting}
\vspace*{-2\baselineskip}
\end{figure}
\section{Applications \& Workloads}
\label{sec:workloads}
Our framework offers a range of workloads to simulate diverse scenarios and stress resource components individually or in combination. These workloads are: (i)~\textit{\textbf{Component-based Stressors}} for targeting specific resources, (ii)~\textit{\textbf{Streaming Analytic Workload}} using popular distributed processing engines, (iii)~\textit{\textbf{ML Inference Application}} with a well-known model and various backends, and (iv)~\textit{\textbf{NoSQL Database Operations}}.
As previously noted, all workloads are containerized and extended to expose application metrics, support parameterization, and enable customization via \textit{Framework Templates}, eliminating the need for code modification or recompilation.

\subsection{Component-based Stressors}
To evaluate individual compute resources or analyze workload co-location, we selected targeted performance tests for each component. For CPU, memory, and disk I/O, we use the~\textit{Linux stress command}~\cite{stress2025}, which applies adjustable stress to components either individually or concurrently. Users can specify 1 to N CPU threads, where N is the CPU’s thread limit, determining the number of workers spawned to simulate different levels of utilization of simple applications.

For network testing, we use~\textit{iperf3}~\cite{iperf2025}
, which transmits data between two devices, acting either as a client or a server. Devices must be on the same local network (wired or wireless) or use public IPs. This workload includes: (i) the \textit{generator}, sending packets at a set rate and measuring outbound traffic, and (ii) the \textit{receiver}, which monitors inbound traffic.



In regards to application-level metrics, one can collect the \textbf{number of packets} exchanged when using the iperf3 tool. As for the stress tool, it does not export any application statistics for direct use, as its purpose is solely to apply stress. 

\subsection{Streaming Analytics Workload}
\begin{figure}[t]
  \centering
     \includegraphics[width=0.43\textwidth] {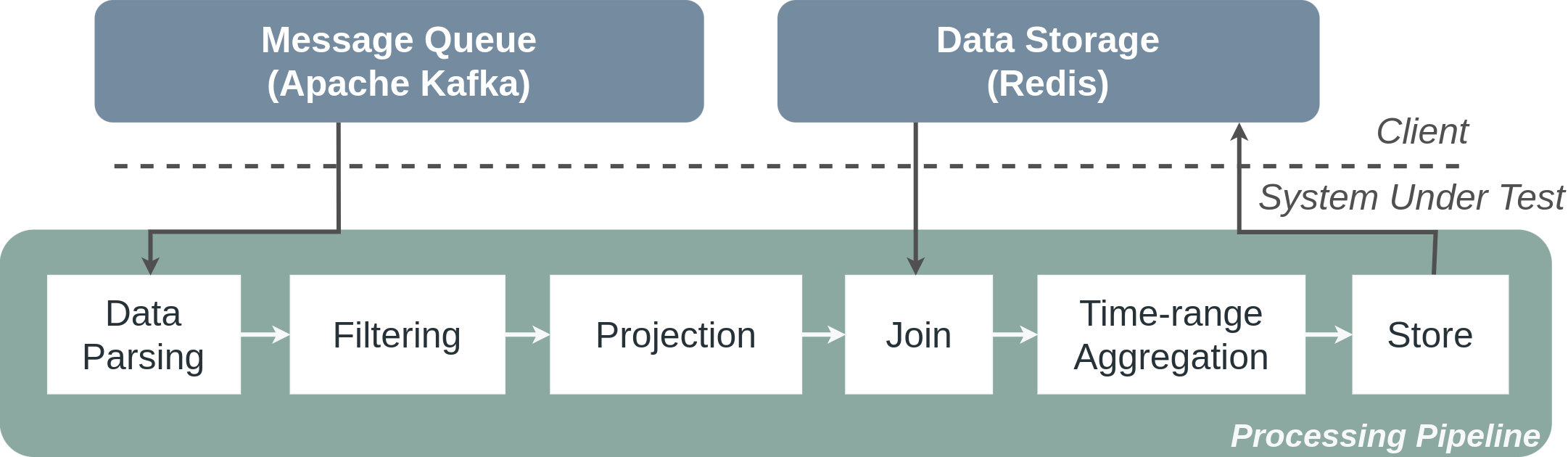}
    \caption{Overview of Yahoo Streaming Workload}
  \label{fig:streaming-workload}
  \vspace*{-1.5\baselineskip}
\end{figure}

For this workload, we have employed the widely known \textit{Yahoo Streaming Benchmark}~\cite{yahooStreaming2016benchmarking}, which is designed to simulate a data processing pipeline for extracting insights from marketing campaigns. As shown in Fig.~\ref{fig:streaming-workload}, the pipeline runs on the edge device, performing tasks such as receiving advertising traffic data, filtering and cleaning it, merging it with existing key-value store information, and storing the final results. 
All data produced by the data generator is pushed and extracted through a message queue (Apache Kafka~\cite{kafka2025}
), while intermediate data and final results are stored in an in-memory database (Redis~\cite{redis2025}
).
This workload can be executed using any of the following distributed streaming processing engines:~\textit{Apache Storm},~\textit{Flink}, or~\textit{Spark}. 
Additionally, we can adjust the data rate, the number of campaigns, workload duration, and engine specific parameters, such as worker partitions, etc. 

To evaluate the performance of this application, we extract two measurements: the \textbf{total number of tuples} processed during execution and the \textbf{total latency} of the application, based on the statistics provided by the selected underlying processing engine for each deployed task.

\subsection{Database Operations}
To compare NoSQL database performance under different conditions, we have utilized the \textit{Yahoo! Cloud Serving Benchmark (YCSB)}~\cite{cooper2010benchmarking}. YCSB applies heavy load to perform basic operations such as reading, updating and inserting records using a single key. It is important to highlight that each run of the workload does not perform only one basic operation per time, but a round of all of them, based on the operation rate that the user defined, until the experimentation time is over. It supports multiple databases, like MongoDB and Redis, and offers three workload distributions: \textit{Zipfian} (frequent access to some items), \textit{Latest} (favoring recent records), and \textit{Uniform} (random access). YCSB also provides adjustable parameters and a flexible schema.
Users can change the number of records to be processed, the total number of operations to be performed, the load distribution, the rate of operations, and the experiment's duration. If all operations are completed before the time limit, the experiment ends early; otherwise, it stops when the time expires. 
Users can increase database stress by specifying the number of threads for asynchronous operations.

YCSB provides statistics for each operation it performs, informing the user of the \textbf{count}, \textbf{min}, \textbf{max}, and \textbf{average} \textbf{operations per second} for each minute of the experiment.


  \vspace*{-.2\baselineskip}
\subsection{Machine Learning Inference Applications}
  \vspace*{-.2\baselineskip}

To evaluate the performance of ML inference tasks, we use the MLPerf~\cite{reddi2020mlperf}, a well-known benchmark for ML training and inference. 
While MLPerf includes numerous tasks, we have focused solely on image classification to create a more targeted workload. This task is widely used in commercial applications to evaluate ML performance. It involves a classifier network that assigns an image to a class. MLPerf uses the ImageNet 2012 dataset, resizing images to 224x224 and measuring Top-1 accuracy. It employs two models: ResNet-50 v1.5, known for high computational demands and strong classification performance, and RetinaNet, effective in object detection with accurate bounding box predictions.

Except for the pre-provided models from MLPerf, we modified its codebase to serve models over the network, enabling measurement of network overhead in ML inference. Specifically, we created two separate services: (i)~a lightweight server that loads models and exposes a RESTful API, and (ii)~a workload generator that loads and sends images to the server one-by-one. We refer to this as ``\textit{streaming mode}", while the default MLPerf setup, which loads images locally, is called ``\textit{local mode}". MLPerf allows users to adjust various workload parameters, including input dataset, max latency, batch size, duration, thread count for asynchronous execution, and the inference framework which is currently limited to CPU-based options: ONNX~\cite{onnx}, NCNN~\cite{ncnn}, or TensorFlow~\cite{tensorflow2015-whitepaper}.


After the workload execution, the retrieved statistics include the model's \textbf{accuracy percentage}, the \textbf{average batches per second} (where each query represents the processing batch of images), the \textbf{total completed queries}, and the \textbf{mean latency}.

\section{Evaluation}
\label{sec:experimentation}


This section comprehensively evaluates our framework's capabilities via experiments based on the use case in Section~\ref{sec:motivation}, demonstrating its ability to assess ML inference services across various configurations and co-located workloads.

\subsection{Experimental Setup \& Devices Under Test}


To realistically replicate MEC nodes and their typical heterogeneity, common across the Multi-access Edge Computing continuum, we selected a range of compute devices under test:


\noindent \textbf{Single-board Computer (SBC)}: A Raspberry Pi 4 Model B with a quad-core ARM Cortex-A72 (4 threads, 1.5GHz) and 4GB RAM. As the most well-known SBC, we chose one of its most popular variants, capable of running multiple workloads.

\noindent \textbf{SMALL server}, that is equipped with a six-core Intel(R) Xeon(R) CPU X5690, which has 12 threads @3.47GHz, 12MB CPU cache, 6.4 GT/s bus speed, and Thermal Design Power (TDP) 130W. Additionally, it comprises a 12GB RAM.

\noindent \textbf{MEDIUM server}, which features a 71GB RAM, and two Intel(R) Xeon(R) CPU E5620 processors. Each CPU has 4 cores and 8 threads at a 2.4GHz frequency, 5.86 GT/s bus speed, 12MB CPU cache, and 80W TDP.

\noindent \textbf{LARGE server}, has 173GB RAM and two Intel(R) Xeon(R) CPU E5-2680v3 CPU processors. Each processor has 12 cores and 24 threads @2.5GHz frequency, 9.6 GT/s bus speed, 30MB cache, and 120W TDP.

All servers feature dual power supply plugs to ensure uninterrupted operation. The \textbf{SMALL}, \textbf{MEDIUM}, and \textbf{LARGE} servers have dynamic frequency scaling (CPU throttling) enabled, a standard Intel CPU feature that improves power efficiency and reduces heat during CPU-intensive workloads.


Lastly, we use a server for (i) the experiments' orchestration, (ii) the deployment of data generators, and (ii) the collection of utilization and application metrics. The latter server has 71 GB RAM, and a 12-core CPU with 24 threads.

\subsection{Workload Parameterization}
For our analysis we have employed the workloads that were previously discussed with specific parameters, namely: 

\noindent \textbf{Component-based stressors}, using them for stressing the: (i)~CPU; (ii)~memory; (iii)~I/O; and (iv)~network.
The primary objective of this workload is to create an intense co-located workload that occupies the device's resources. 
In all stress scenarios, the number of workers assigned to each device was based on the maximum number of threads supported. 
We spawned 4 workers for the Raspberry Pi, 12 for the small server, 16 for the medium server, and 48 for the large server.

\noindent \textbf{Streaming workload} focuses on Apache Storm as the streaming processing engine, even though our experiments could be easily done using any of the other two engines, Spark and Flink. In regards to workload parameters, we have used 1000M tuples per second data generation rate and 100M campaigns. 

\noindent \textbf{Database workload} employs MongoDB as the underlying database, using a Zipfian load distribution and the default YCSB workload configuration values, with two exceptions: the number of records and the number of operations, which for both we used a value of 2.5M. Additionally, we chose to run this benchmark in asynchronous mode, using 12 threads, in order to stress our cluster to its maximum capacity. 


\begin{figure*}[!t]

  \centering
     \includegraphics[width=.98\textwidth, trim=25 0 25 8, clip] {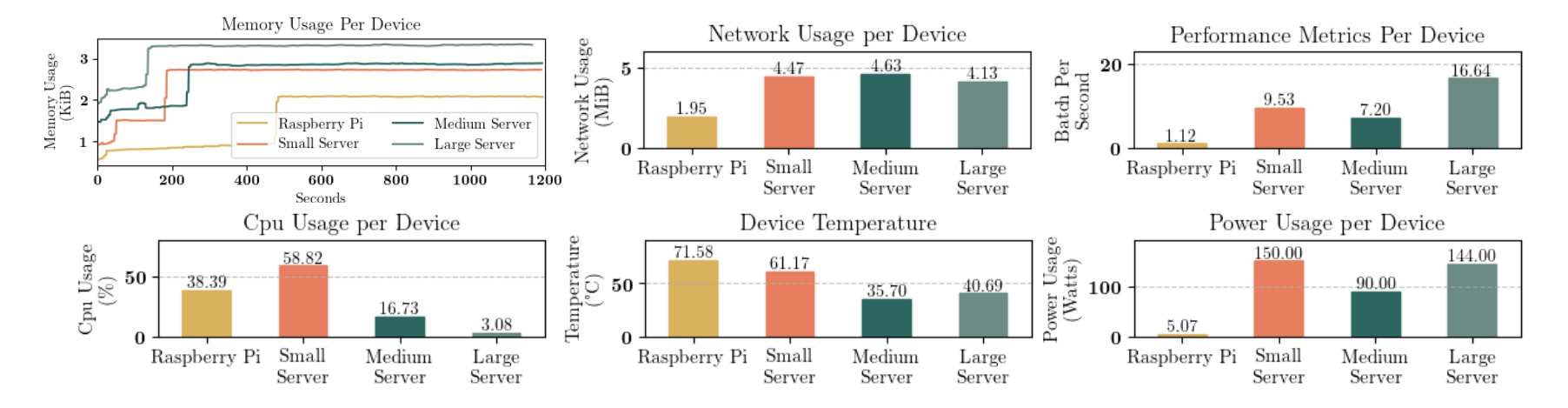}
  \vspace*{-1\baselineskip}

    \caption{Utilization and Performance Metrics Across Devices Using an ML Inference Application}
  \label{fig:usecase1}
  \vspace*{-1.5\baselineskip}
  \end{figure*}

\noindent \textbf{ML workload} uses the image classification for both local and streaming mode, utilizing the following inference frameworks: ONNX, NCNN, and TensorFlow. Again, we deployed this workload using asynchronous processing with 12 threads.

For all the aforementioned workloads, \textit{the selection of parameters was made based on trial and error}, where we tested a variety of values until we identified the silver lining of load to stress our cluster without overwhelming it. 

  \vspace*{-.5\baselineskip}
\subsection{Results and Discussion}
In this section, we explore three key research questions related to the use case scenario outlined in Section~\ref{sec:motivation}. These questions not only demonstrate the usability of our system but also provide valuable insights into service co-location on the edge, benefiting both researchers and practitioners.
All scenarios are executed for 20 minutes utilizing our framework and capturing utilization and application-level metrics. Additionally, all of our utilization plots represent the mean value over the entire execution (e.g., power consumption, measured in watts, represents the average rate of power usage).

\medskip

\textit{\textbf{RQ1: Given an edge cluster, which device is best suited for execution of a streaming ML inference application in terms of performance, resource utilization and energy efficiency?}}

Let us consider a scenario, where a group of administrators have at their disposal a set of unused devices and want to deploy a streaming ML inference application. So, they are interested in evaluating which device performs best with this workload. To replicate this scenario, we deployed the ML inference application in streaming mode across all of the devices under~test. It is important to highlight that the workload deployment was chosen to be isolated at this stage, allowing it to serve later as a baseline for comparison.

The first plot (upper-left corner of Fig.~\ref{fig:usecase1}) shows the memory usage for the different nodes that follow a similar pattern.
Initially, memory utilization is low across all nodes since the model has not yet been loaded. At a certain point, we notice a distinct "step", where the model occupies a significant portion of memory, maintaining this level until the end.
However, the delay in memory allocation (cold start effect) differs depending on the machine. The \textit{Large Server} loads the model fastest, followed by the \textit{Small} and \textit{Medium Servers}, while the \textit{Raspberry Pi} is the slowest. Additionally, baseline memory usage and the devices' storage differences also impact the loading time. To ensure fair analysis, the results in the following plots and sections consider only the period after model loading.

Next, we examined CPU usage, power needs, and temperature metrics (Fig.~\ref{fig:usecase1}, second series of plots), as these factors are closely correlated.
We observe that both the \textit{Raspberry Pi} and \textit{Small Server} have a higher CPU usage, contrary to the \textit{Medium} and \textit{Large Servers}, where the CPUs remain underutilized.
Moreover, the \textit{Small Server} has the highest power demand during inference, consuming 150 watts, followed by the \textit{Large Server} (144 watts) and the \textit{Medium Server} (90 watts). The \textit{Raspberry Pi}, due to its low-power profile, requires only 5 watts.
However, when analyzing temperature, the \textit{Raspberry Pi} reaches the highest levels at 71.5$^\circ$C, followed by the Small Server at 61$^\circ$C. This indicates that both devices are operating under significant load. In contrast, the \textit{Medium} and \textit{Large Servers} maintain lower temperatures (35$^\circ$C - 40$^\circ$C).

Moreover, we examine the performance and the overall network traffic of each device (Fig.~\ref{fig:usecase1} first line middle and right plots).
Starting with throughput (processed batch of images per second), \textit{Raspberry Pi} exhibits the lowest performance by a significant margin, processing only 1.12 batches per second, while \textit{Small}, \textit{Medium}, and \textit{Large Servers} process 9.5, 7.2, and 16.6 batches per second, respectively.
Interestingly, \textit{Small Server} achieves higher throughput than \textit{Medium Server}, but the latter offers a better balance between performance and CPU utilization, resulting in lower power consumption.
In network traffic, due to its limited processing throughput, the \textit{Raspberry Pi} generates approximately half of the network traffic of the other deployments. 
For the remaining nodes, data transfer remains relatively consistent, with minor differences related to the randomness of batching images before transmission.

Finally, we evaluated the energy efficiency of each node by calculating the average energy consumption per processed batch.
Our results show that the \textit{Raspberry Pi} is the most energy-efficient node, requiring only 4.52 joules per batch. Among the servers, the \textit{Large Server} performs best, consuming 8.65 joules per batch, followed by the \textit{Medium Server} and the \textit{Small Server} at 12.5 and 15.73 joules per batch, respectively.

\noindent \textit{\textbf{Key Takeaways:}} 
To this end, small edge devices (e.g., Raspberry Pi) seem to have lower performance in model loading and throughput, but are the most energy-efficient options (6.67J per batch). 
Small servers are not always energy efficient and may have temperature issues, yet may match or surpass the medium servers in throughput. 
Large servers deliver the best performance and may provide energy efficiency but has high static power demands, while medium servers balance power needs, energy efficiency and throughput.
\smallskip




\textit{\textbf{RQ2: Given an edge device which already has deployed a streaming ML inference application, how can other different co-located services affect its performance?}}

After evaluating the first scenario, the administrator has chosen to deploy the ML inference algorithm on the \textit{Medium Server}. However, with no other servers available, the user must co-locate the ML workload with other tasks, such as component-intensive stressors (e.g., CPU, memory, disk, and network) or general-purpose workloads (e.g., databases or streaming analytics). To determine whether this setup is feasible and how it impacts performance, the user leverages our framework again, modifying only the input parameters.

Our first observation is that cold start duration varies across different co-location strategies.
Specifically, when deploying the ML streaming inference workload alongside a network stressor (iperf), the impact on cold start time is minimal (less than 10\%). In contrast, collocating the service with CPU and disk I/O stressors increases cold start duration by approximately 35-40~\%. Most notably, introducing a memory-intensive stressor doubles the cold start time, rising from 244 seconds to 505 seconds. This underscores the significant impact of memory pressure on ML model initialization.

\begin{figure}[!t]
  \centering
     \includegraphics[width=0.48\textwidth, trim=25 0 25 22, clip] {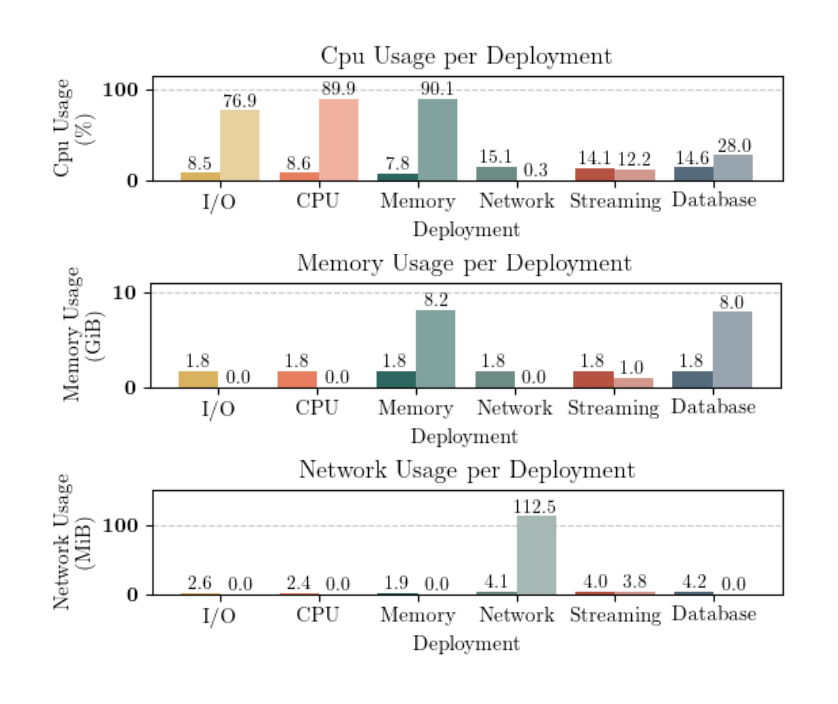}
       \vspace*{-1.5\baselineskip}

    \caption{CPU, memory \& network usage during co-location of an ML inference application with other workloads. Darker shades represent the ML workload's resource consumption, while lighter indicate the co-located workload's usage.}
  \label{fig:cpu-mem-disk}
  \vspace*{-1.5\baselineskip}
  \end{figure}

Moreover, Fig.~\ref{fig:cpu-mem-disk} (first plot) illustrates the CPU utilization of each workload across different experiments.  
Our key observation is that component-based stressors significantly reduce the CPU utilization of the ML streaming service. In an isolated execution, the service utilizes 16.73\% of the CPU (Fig.~\ref{fig:usecase1}, right-bottom plot). However, when co-located with disk I/O, CPU, and memory stressors, its utilization drops to approximately 8\%, as these stressors consume a substantial share of CPU resources: 76.82\% (disk I/O), 89.93\% (CPU), and 90.10\% (memory).  
When the co-located workloads require less CPU, such as the network stressor (0.29\%), database (27.98\%), and streaming analytics (14.14\%), the CPU allocation for the ML workload nearly doubles to approximately 14-15\%, which is closer to the 17.67\% of isolated execution scenario.

Regarding memory and network utilization (Fig.~\ref{fig:cpu-mem-disk}, last two plots), we observe minimal impact on the corresponding metrics of the target workload (ML streaming).  
Interestingly, when workloads introduce network stress, such as the network stressor or streaming analytics, the target workload generates more network traffic. Although this may seem counterintuitive, it can be explained by considering CPU utilization and throughput. Specifically, we observe that more data points are processed and transmitted over the network when the CPU usage of the co-located workloads is low.  
Moreover, the memory allocation of the memory stressor and the database appears to have minimal impact on the target workload, except for the cold start period, as previously discussed. 

\begin{figure}[!t]
  \centering
     \includegraphics[width=0.48\textwidth, trim=25 0 25 22, clip] {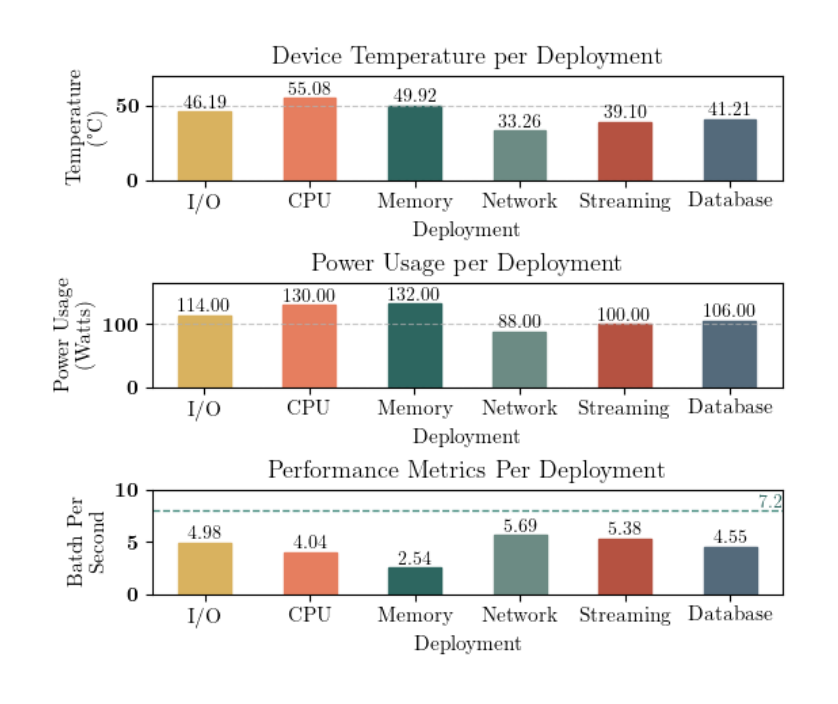}
       \vspace*{-1.7\baselineskip}
    \caption{Temperature, Power Usage, and Performance Metrics for co-located workloads (ML+Workload)}
  \label{fig:temperature-power-disk}
  \vspace*{-1.5\baselineskip}
  \end{figure}

Additionally, first and second plots of Fig.~\ref{fig:temperature-power-disk} show the server's temperature and power consumption, respectively.  
While workload co-location inevitably increases both metrics, different workload types contribute to varying levels of impact.  
For instance, CPU-intensive stressors significantly elevate energy demand, with power consumption rising from 88 watts (baseline) to 114 watts, 130 watts, and 132 watts for I/O, CPU, and memory stressors, respectively.  
In contrast, real-world workloads, such as database operations (105 watts) and streaming analytics (100 watts) lead to more moderate increases.  
Moreover, power consumption and temperature exhibit a strong correlation, with the highest temperature observed under CPU stress, followed by memory and I/O stressors. Notably, network stress has negligible effects on power and temperature, reevaluating the relationship between CPU usage and system heat generation.  

Finally, last plot of Fig.~\ref{fig:temperature-power-disk} illustrates the throughput (Batches per second) of the ML streaming workload and the impact of service co-location.
The green dashed line represents the throughput without co-location (7.29 batches per second). When co-located with disk I/O, network, database, and streaming workloads, performance degradation occurs, reducing throughput to approximately 4.5--5.5 Batches per second.
The workloads that affect the ML application are the CPU-intensive workload and the memory-intensive workload.
Examining the utilization metrics in Fig.~\ref{fig:cpu-mem-disk}, we observe that the CPU-intensive workload primarily consumes CPU resources, while the memory stressor heavily utilizes both CPU and memory. 
The latter has a greater negative impact on the ML workload, reducing its throughput to 2.54 Batches per second, whereas the CPU-intensive workload allows for a higher throughput of 4.04 Batches per second.
The plot highlights a strong correlation between available CPU resources and the ML streaming service's throughput, as well as the influence of other compute resources, such as memory.  

\noindent \textbf{\textit{Key Takeaways: }}
In this scenario, we highlighted that memory-intensive stressors significantly increase the cold start time of ML services, which is important to consider, especially when there is a need to restart the workload frequently. Furthermore, CPU and disk I/O stressors substantially reduce the CPU allocation for the ML workload, leading to degraded performance. 
Additionally, when co-located workloads demand less CPU, the ML service retains higher CPU utilization, improving throughput, while memory and network stressors have minimal direct impact. 
Moreover, CPU-intensive stressors significantly increase power consumption and temperature, while real-world workloads have a more moderate impact, and network stressors exhibit negligible effects, reinforcing the strong correlation between CPU usage and system heat~generation.


  \smallskip
\textit{\textbf{RQ3: Given an Edge device which already has deployed multiple workloads, like ML, databases, and streaming analytics, how can different configurations of them affect the overall performance and utilization metrics?}}

In this session, we examine a scenario where the user has already deployed workloads on a selected node (\textit{Medium Server}) and intends to execute an ML workload alongside them. The user also aims to compare different execution backends in various modes to find the best option for throughput and energy efficiency, while also identifying any potential uncharted trade-offs.
For this, we co-locate the ML workload with streaming analytics and database workloads while also modifying the benchmarking configuration of the ML workload. We assess three different execution engines (backends) -- ONNX, NCNN, and TensorFlow (TF) -- under two execution modes: (i)~streaming (remote), where image batches are sent over the network from a workload generator to the stressed node; and (ii)~local execution, where images exist on the same machine, allowing inference without network involvement. 

Examining the cold start period across different deployments, we observed that the ONNX and NCNN engines load the model significantly faster (30--40 seconds) compared to TensorFlow (356--375 seconds) in all cases. So, specialized ML engines optimized for CPU compute resources can substantially reduce the startup time in such deployments. Moreover, we have to note that the variations on co-located services (at least for database and streaming analytic process) do not influence the starting period of ML workloads.

\begin{figure}[!t]
  \centering
     \includegraphics[width=0.48\textwidth, trim=25 0 26 20, clip] {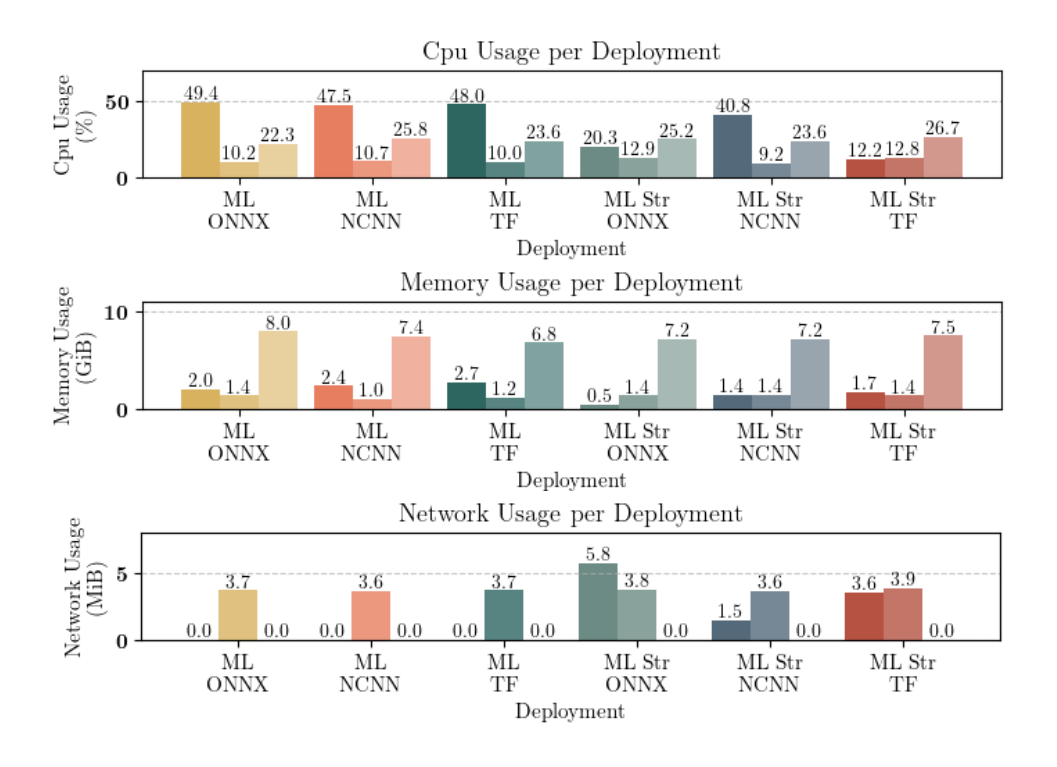}
       \vspace*{-1.4\baselineskip}

    \caption{CPU, memory, and network usage of an ML inference application co-located with other workloads. In each group, the first shade represents the ML workload, the middle the Streaming Analytic workload, and the right the Database’s resource consumption.}
  \label{fig:cpu-mem-disk-co-location}
  \vspace*{-1.5\baselineskip}
  \end{figure}

Fig.~\ref{fig:cpu-mem-disk-co-location} depicts the utilization of CPU, memory, and network resources when an ML inference application runs alongside other workloads. In each group, the first shaded section (left bar) represents the ML workload, the middle one corresponds to the Streaming Analytics workload, and the rightmost bar indicates the resource consumption of the Database workload.
Firstly, we observe that local execution (ML ONNX~/~NCNN~/~TF) leads to higher CPU utilization for ML workloads, reaching just below 50\%, while the database and streaming analytics workloads remain at approximately 23\% and 10\%, respectively. In contrast, when the ML workload processes batches of images transmitted over the network (ML Str ONNX~/~NCNN~/~TF), the CPU utilization decreases to 20.3\% for ONNX, 40.8\% for NCNN, and 12.2\% for TF.  
This reduction is primarily due to network-related operations (e.g., encoding/decoding, HTTP handshake, etc.) handled by the network connector, during which the CPU remains idle. As a result, more CPU capacity becomes available for the co-located services, leading to a slight increase in their CPU utilization, ranging from 23.6\% to 26.7\% for the database workload and from 9.2\% to 12.8\% for streaming analytics.

  \begin{figure}[!t]
  \centering
     \includegraphics[width=0.48\textwidth, trim=25 0 25 23, clip] {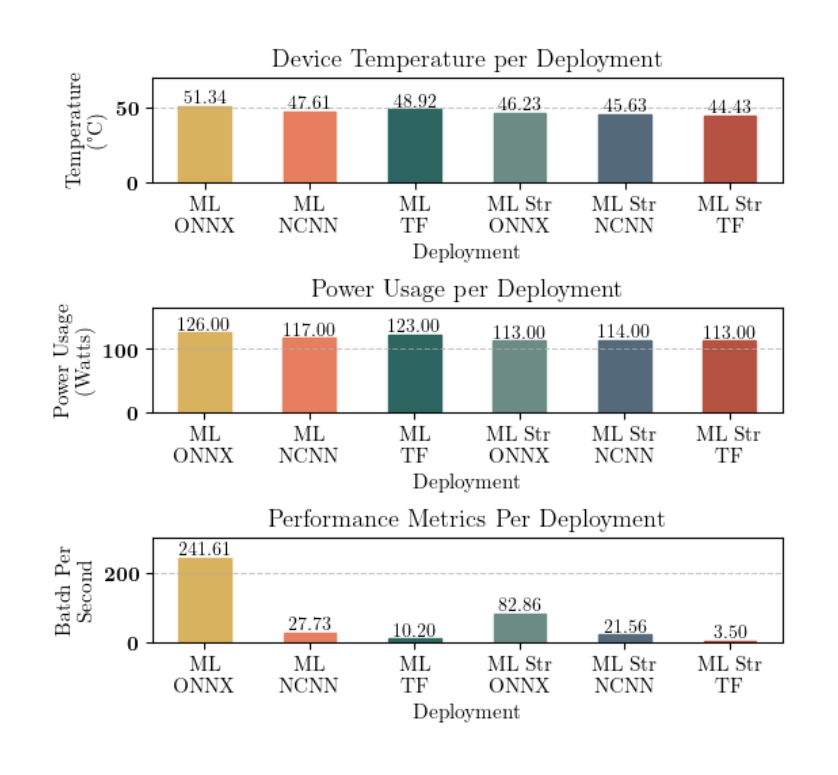}
       \vspace*{-1.6\baselineskip}
    \caption{Temperature, Power Usage, and Performance Metrics for co-located workloads (ML+Streaming+Database)}
  \label{fig:temperature-power-co-location}
  \vspace*{-1.5\baselineskip}
  \end{figure}

For memory usage, we saw almost no differences in the database and streaming analytics workloads across execution modes, with the database workload consuming the largest portion of the server's memory (7--8GB).  
In contrast, the ML shows differences in memory allocation between local (2--3GB) and remote execution (1.4--1.7GB), as in local execution, the workload loads the entire dataset into memory, whereas in remote execution, it only fetches the model's weights.

Regarding network usage, streaming analytics dominate, generating approximately 3.7 MB of network traffic during the experiment. Neither database workload nor local ML execution involve network data transfers.  
Interestingly, ML streaming (remote) workloads consume a significant amount of network resources, with ML Str ONNX surpassing the streaming analytics workload and ML Str TF showing similar network usage.  
To understand this behavior, one must consider the rate and size of the input data for streaming workloads (both analytics and ML). Streaming analytics handle numeric data at a higher throughput than ML services, whereas ML workloads process batches of images. Although images are larger in size compared to numeric data, their processing takes longer, usually resulting in lower throughput. 
However, since ONNX achieves a higher processing rate, it requires more network bandwidth compared to the streaming analytics~workload.

Fig.~\ref{fig:temperature-power-co-location} presents the server's temperature, power consumption, and the performance metrics of the ML workload.  
The temperature during local ML execution (ML~ONNX~/~NCNN~/~TF) is slightly higher, ranging from 47.61$^{\circ}$C to 51.19$^{\circ}$C, compared to streaming ML execution (ML~Str~ONNX~/~NCNN~/~TF), which varies between 44.43$^{\circ}$C and 46.23$^{\circ}$C.  
Moreover, local execution requires more power, consuming between 117--127 watts, whereas remote execution operates with a lower power demand of 113--114 watts.

The last plot of Fig.~\ref{fig:temperature-power-co-location} shows the throughput (batch per second) of ML services.
In all cases, local ML execution provides much better results than the same workload running remotely. Specifically, the best performance is observed for local ONNX with 241.61 batches/second. The second position is also allocated by the ONNX but for its streaming execution with 82.86 batches/second. The latter highlights the superiority of ONNX execution engine for CPU inference. 
NCNN handles 27.73 batches/second and 21.56 batches/second, for local and remote execution, respectively, while the worst performance observed for TensorFlow (10.2 and 3.5 batches/second). 

\noindent \textit{\textbf{Key Takeaways:}}  
Our analysis highlights that ONNX and NCNN achieve significantly faster cold start times than TensorFlow, due to the benefits of CPU-optimized ML engines, which is particularly advantageous when someone wants to change models frequently.   
Local execution results in higher CPU usage but reduces network traffic, while remote execution lowers ML CPU usage, allowing co-located services to utilize more resources. 
ML workloads require more memory during local execution (2--3GB) than remote (1.4--1.7GB), and the latter workloads consume notable network bandwidth, with ONNX demanding the most due to its high throughput.  
Local ML execution leads to slightly higher temperatures (47.61$^{\circ}$C--51.19$^{\circ}$C) and power consumption (117--127W) than remote execution (44.43$^{\circ}$C--46.23$^{\circ}$C, 113--114W).  
Performance-wise, ONNX outperforms other backends, achieving the highest throughput (241.61 batches/sec locally, 82.86 remotely), while TF performs the worst (10.2 and 3.5 batches/sec).

\section{Conclusion}
\label{sec:conclusion}

In this work, we introduced an automated benchmarking framework 
to evaluate the impact of multi-tenancy in Edge Computing environments. 
It simplifies the deployment, monitoring, and analysis of co-located workloads, enabling reproducible performance evaluation across diverse hardware setups.
Through a series of experiments, we showed how workload co-location impacts key performance metrics, including cold start duration, CPU utilization, and network behavior. 
Specifically, our \textit{key findings} include: (i) small edge devices, while energy-efficient, have slower model loading and throughput compared to larger servers, which offer the best performance but come with significantly higher power demands; (ii) memory-intensive workloads increase the cold start time of machine learning applications by nearly 1.5x, while CPU and disk I/O stressors degrade performance through heavy CPU usage; and (iii) CPU-optimized ML engines, such as ONNX and NCNN, achieve significantly faster cold starts (up to 11×), with ONNX achieving the highest throughput, surpassing NCNN by 9x and TensorFlow by 24x. 
Building on these insights, our tool allows researchers to generate profiles of utilization metrics such as memory, CPU, disk, and energy, for each 
workload combination and configuration.
These profiles are valuable for co-location-aware strategies, while practitioners can use the benchmarking loops to monitor and adapt placement policies based on specific target metrics.

For \textbf{future work}, we plan to evaluate the impact of the multitenancy of modern ML workloads, deployed on GPUs or edge accelerators. 
Additionally, we will extend the framework to support the creation of customizable edge network topologies, 
by using programmable network devices or emulators, in order for our framework to be capable of changing not only workload parameters but also infrastructure configurations. 
Most importantly, we aim to enhance the framework with the capability to automatically identify optimal workload parameters and configurations, apply them, and allocate them to edge nodes.
To this end, we will explore automated methods, such as Bayesian optimization, ML techniques, and reinforcement learning, to determine 
deployment parameters based on specified performance indicators, including energy consumption, latency, and computational footprint.


\smallskip
\footnotesize{\noindent \textbf{Acknowledgement.} This work is part of AdaptoFlow that has indirectly received funding from the European Union’s Horizon Europe research and innovation action programme, via the TRIALSNET Open Call issued and executed under the TrialsNet project (Grant Agreement no. 101017141).}
\bibliographystyle{IEEEtran}
\bibliography{main.bib}

\vfill
\end{document}